\documentclass{IEEEtran}
\usepackage{cite}
\usepackage{amsmath,amssymb,amsfonts}
\usepackage{graphicx}
\usepackage{textcomp,nicefrac}
\def\BibTeX{{\rm B\kern-.05em{\sc i\kern-.025em b}\kern-.08em
T\kern-.1667em\lower.7ex\hbox{E}\kern-.125emX}}
\begin{document}
\title{The FragmentatiOn Of Target Experiment (FOOT) and its DAQ system}
\author{Silvia Biondi, Andrey Alexandrov, Behcet Alpat, Giovanni Ambrosi, Stefano Argir\`o, Rau Arteche Diaz, Nazarm Bartosik, Giuseppe Battistoni, Nicola Belcari, Elettra Bellinzona, Maria Giuseppina Bisogni, Graziano Bruni, Pietro Carra, Piergiorgio Cerello, Esther Ciarrocchi, Alberto Clozza, Sofia Colombi, Giovanni De~Lellis, Alberto Del~Guerra, Micol De~Simoni, Antonia Di~Crescenzo, Benedetto Di~Ruzza, Marco Donetti, Yunsheng Dong, Marco Durante, Veronica Ferrero, Emanuele Fiandrini, Christian Finck, Elisa Fiorina, Marta Fischetti, Marco Francesconi, Matteo Franchini, Gaia Franciosini, Giuliana Galati, Luca Galli, Valerio Gentile, Giuseppe Giraudo, Ronja Hetzel, Enzo Iarocci, Maria Ionica, Keida Kanxheri, Aafke Christine Kraan, Chiara La~Tessa, Martina  Laurenza, Adele Lauria, Ernesto Lopez Torres, Michela Marafini, Cristian Massimi, Ilaria Mattei, Alberto Mengarelli, Andrea Moggi, Maria Cristina Montesi, Maria Cristina Morone, Matteo Morrocchi, Silvia Muraro, Livio Narici, Alessandra Pastore, Nadia Pastrone, Vincenzo Patera, Francesco Pennazio, Pisana Placidi, Marco Pullia, Fabrizio Raffaelli, Luciano Ramello, Riccardo Ridolfi, Valeria Rosso, Claudio Sanelli, Alessio Sarti, Gabriella Sartorelli, Osamu Sato, Simone Savazzi, Lorenzo Scavarda, Angelo Schiavi, Christoph Schuy, Emanuele Scifoni, Adalberto Sciubba, Alexandre S\'echer, Marco Selvi, Leonello Servoli, Gianluigi Silvestre, Mario Sitta, Eleuterio Spiriti, Giancarlo Sportelli, Achim Stahl, Sandro Tomassini, Francesco Tommasino, Marco Toppi, Giacomo Traini, Tioukov Valeri, Serena Marta Valle, Marie Vanstalle, Ulrich Weber, Antonio Zoccoli, Roberto Spighi, and Mauro Villa
\thanks{S. Biondi is with the University ``Alma Mater Studiorum'' of Bologna and Instituto Nazionale di Fisica Nucleare (INFN, National Istitute of Nuclear Physics), Italy (e-mail: silvia.biondi@cern.ch - silvia.biondi@bo.infn.it).}}

\maketitle

\begin{abstract}
The FragmentatiOn Of Target (FOOT) experiment aims to provide precise nuclear cross-section measurements for two different fields: hadrontherapy and radio-protection in space. The main reason is the important role the nuclear fragmentation process plays in both fields, where the health risks caused by radiation are very similar and mainly attributable to the fragmentation process.
The FOOT experiment has been developed in such a way that the experimental setup is easily movable and fits the space limitations of the experimental and treatment rooms available in hadrontherapy treatment centers, where most of the data takings are carried out. The Trigger and Data Acquisition system needs to follow the same criteria and it should work in different laboratories and in different conditions. It has been designed to acquire the largest sample size with high accuracy in a controlled and online-monitored environment. The data collected are processed in real time for quality assessment and are available to the DAQ crew and detector experts during data taking.
\end{abstract}


\section{Introduction}
\label{sec:introduction}
\IEEEPARstart{T}{he} main goal of the FOOT experiment is to provide nuclear cross-section measurements necessary in two different fields: hadrontherapy and radio-protection in space.
\\In the last decade, a continuous increase in the number of cancer patients treated with Particle Therapy (PT) had been registered, due to its effectiveness in the treatment of deep-seated solid tumors which cannot be treated with surgery \cite{b1}. When the charged particles travel through the patient, nuclear interactions occur producing nuclear fragments that can cause side effects in regions outside the tumor volume. As a consequence a precise evaluation of this effect, at the hadrontherapy energies (150--400 MeV/$u$), would increase the accuracy of the treatment.
\\Regarding to the second FOOT field of interest, the XXI century will be characterized by a deeper exploration of the Solar System that will involve long term human missions as the expedition to Mars. Health risks are associated to exposure to Galactic Cosmic Rays (GCR), that are very energetic (on average around 700--1000 MeV/$u$) and produce showers of light fragments and neutrons by nuclear fragmentation when hitting the spaceship shields. Considering that the GCR are composed of 90\% of protons, 9\% of helium and the rest of heavy nuclei, the overlap with the measurements for hadrontherapy purposes is large, the main difference being the energy range.
\\Regarding physical parameters, target fragmentation plays a key role as low energy secondary fragments contribute to increment the dose deposition in human body tissues along the entrance channel, in case of both PT and radio-protection in space, and in the region surrounding the tumor, in case of PT treatment.
The complexity of dedicated experiments makes the evaluation of the secondary fragmentation challenging, and in fact very few and limited experimental data are available in literature regarding target fragmentation, and none of them fully describes secondary fragments induced by a charged particles beam.
\\In this scenario, the FOOT collaboration, made of about one hundred physicists from France, Germany, Italy and Japan, has the purpose to perform precise measurements of differential cross sections, with respect of the kinetic energy and the production angle of the emitted fragments. The FOOT measurements campaign foresees an extensive program focused on the nuclear fragmentation of $^4$He, $^{12}$C and $^{16}$O beams impinging on thin C and hydrogen rich targets, like polyethylene C$_2$H$_4$, in the energy range 100$-$800 MeV/nucleon, of interest for PT and radioprotection in space.
\\In this paper, the experimental setup is described in detail and the Trigger and Data Acquisition system is reported and discussed in all its aspects.

\section{The FOOT experiment}
\label{sec:footExp}
The FOOT experiment has been designed to detect, track and identify all the charged fragments produced in ion collisions with different targets, with the aim of measuring both projectile and target fragmentation. The latter, which is of interest for applications in the proton-Nucleus (p-N) collisions field, is a very challenging task, due to the short range of the produced fragments, resulting in a very low probability of escaping the target. Indeed, the before-mentioned range is less than tens of micrometers, thus, even a very thin solid target would stop them or badly spoil their energy measurement. In this experiment, the inverse kinematic approach is used in order to overcome this issue: fragmentations of different ion beams (like $^{12}$C and $^{16}$O) onto hydrogen enriched targets, at 50-200 MeV/nucleon will be studied.
In this case, the p-N cross sections should be extracted by subtraction from data taken on C$_2$H$_4$ and C targets, as explained and discussed in \cite{b2} and \cite{b3}.

Since FOOT has a detector capable of performing the target fragmentation measurement with the approach mentioned above, it can as well perform the direct measurement of projectile fragmentation cross sections, induced by C, He and O beams on the same graphite and polyethylene targets, for PT application, and explore the higher incoming beam energy range, for applications to the radio protection in space. Using different target materials, it can study collisions with other nuclei of interest for biological effects as well. Table~\ref{tab1} and \ref{tab2} report the physics programm of the FOOT experiment, for the PT and radio protection in space applications respectively, where PMMA is polymethyl methacrylate (C$_5$O$_2$H$_8$)$_\mathrm{n}$.

\begin{table}
	\label{table1}
	\caption{FOOT physics program for the hadrontherapy application.}
	\setlength{\tabcolsep}{3pt}
	\begin{tabular}{p{75pt}|p{25pt}|p{40pt}|p{30pt}|p{35pt}}
		\hline
		Physics & Beam & Target & Upper Energy (MeV/$u$) & Kinematic approach \\
		\hline
		Target fragmentation & $^{12}$C & C, C$_2$H$_4$ & 200 & inverse\\
		Target fragmentation & $^{16}$O & C, C$_2$H$_4$ & 200 & inverse\\
		\hline
		Beam fragmentation & $^4$He & C, C$_2$H$_4$, PMMA & 250 & direct\\
		Beam fragmentation & $^{12}$C & C, C$_2$H$_4$, PMMA & 400 & direct\\
		Beam fragmentation & $^{16}$O & C, C$_2$H$_4$, PMMA & 500 & direct\\
		\hline
	\end{tabular}
	\label{tab1}
\end{table}

\begin{table}
	\label{table2}
	\caption{FOOT physics program for the radio protection in space application.}
	\setlength{\tabcolsep}{3pt}
	\begin{tabular}{p{75pt}|p{25pt}|p{40pt}|p{30pt}|p{35pt}}
		\hline
		Physics & Beam & Target & Upper Energy (MeV/$u$) & Kinematic approach \\
		\hline
		Beam fragmentation & $^4$He & C, C$_2$H$_4$, PMMA & 800 & direct\\
		Beam fragmentation & $^{12}$C & C, C$_2$H$_4$, PMMA & 800 & direct\\
		Beam fragmentation & $^{16}$O & C, C$_2$H$_4$, PMMA & 800 & direct\\
		\hline
	\end{tabular}
	\label{tab2}
\end{table}

The main goal of the FOOT experiment is to measure differential cross sections with respect to the kinetic energy ($\mathrm{d}\sigma/\mathrm{d}E_{\mathrm{kin}}$) for the target fragmentation process with a precision better than 10\% and double differential cross sections ($\mathrm{d}^2\sigma/\mathrm{d}E_{\mathrm{kin}}/ \mathrm{d}\theta$) for the projectile fragmentation process with a precision better than 5\%. This requires a capability of fragments charge and isotopic identification at the level of 2--3\% and 5\%, respectively, in order to have a clear separation of all the nuclides under study.
These requirements make the measurement particularly challenging with the inverse kinematic approach. In this case, the momentum and kinetic energy have to be measured with a resolution of the order of percent and the nuclide emission angle with respect to the beam direction have to be determined with a resolution of few milliradiants. 
All these aspects have been taken carefully into account when designing the FOOT experimental setup, in terms of target, sub-detectors, structure and total size of the experiment.

\section{Experimental setup}
\label{sec:exp_apparatus}
The target thickness of the FOOT experiment has been designed to be of the order of 2--4 g/cm$^2$, in order to minimize the multiple scattering process and the probability of secondary fragmentation inside the material, and subsequently to reach the resolutions needed for the measurements planned in the physics program.

An important aspect enetering the design of the FOOT experiment is the fact that it needs to be a ``movable'' detector in order to fit the space limitations set by the possible experimental rooms where ion beams are available at hadrontherapy energies. This leads to design an experimental setup with as much as possible limited length and weight.

\begin{figure}[t]
	\centerline{\includegraphics[width=3.5in]{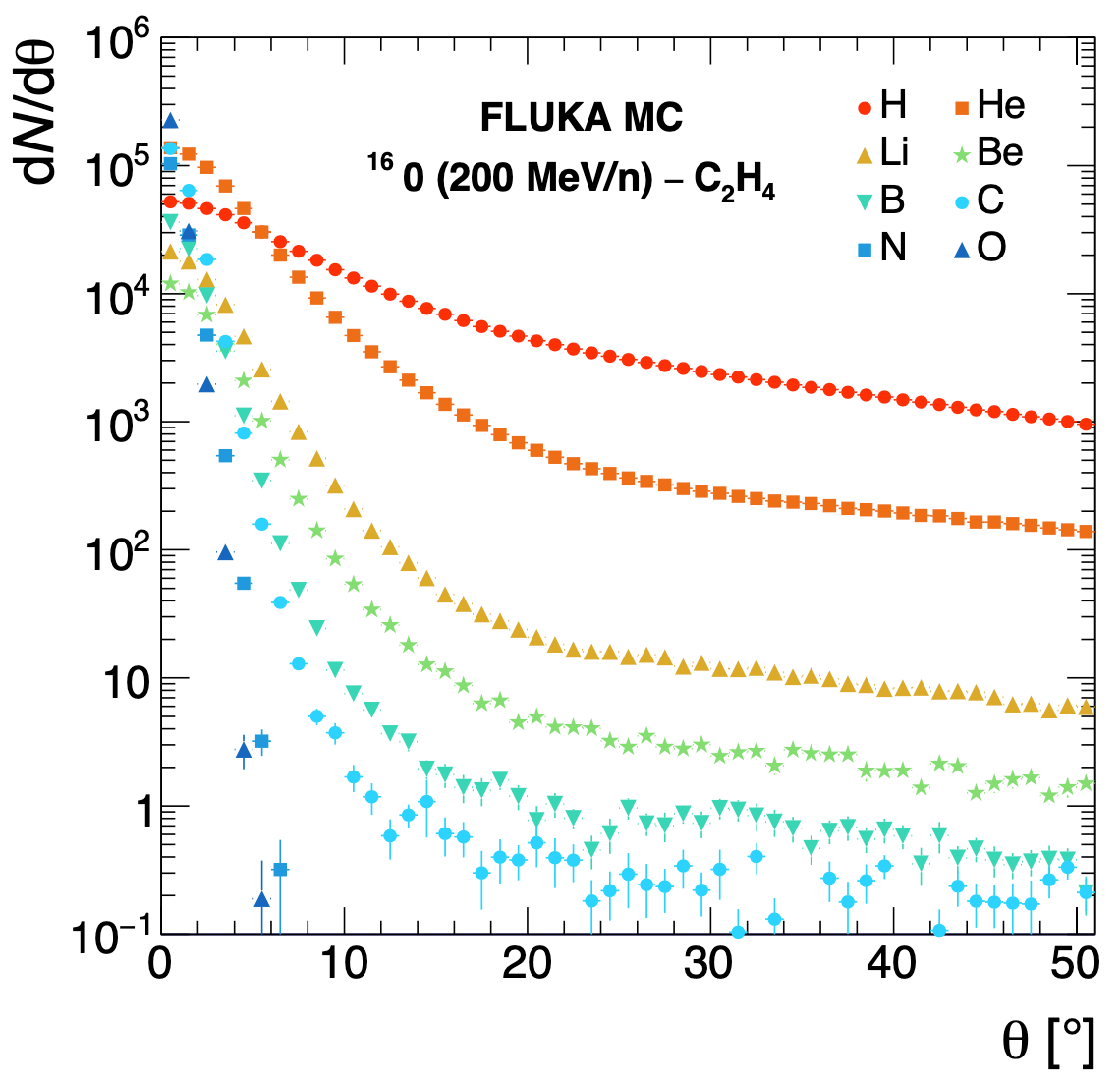}}
	\caption{MC simulation of the angular distributions of different fragments produced by a 200 MeV/nucleon $^{16}$O beam impinging on a C$_2$H$_4$ target.}
	\label{fig_angle}
\end{figure}

Moreover, nuclear fragmentation produces both light and heavy fragments: the first are produced within a wide opening angle, while the second close to the beam direction. It can be seen in Fig.\ref{fig_angle},where the simulations (using FLUKA \cite{b4} ) in terms of fragments yields as a function of the emission angle show that heavier fragments (with a charge Z higher than 2) populate the lower values range of emission angle, below 10$^\circ$, while the light fragments have a wider angular distribution. This behaviour has been considered in the FOOT detector acceptance design.

In order to detect both types of fragments and to fit the design constraints of a ``table top'' setup discussed above, the FOOT detector consists of two different and complementary configurations: an electronic and an emulsion chamber setup.

\begin{enumerate}
\item a setup based on a magnetic spectrometer, together with detectors for tracking sand others optimized for the identification of heavier fragments (Z$\geq$4), This setup has an angular coverage up to a polar angle of $\sim$10$^\circ$ with respect to the beam direction;
\item a setup based on an emulsion spectrometer, optimized for lighter fragments (Z$<$4) identification with an angular coverage larger with respect to the magnetic spectrometer one, extending it upd to $\sim$70$^\circ$.
\end{enumerate}

\subsection{Electronic Setup}
\label{subsec:electronic}
\begin{figure}[t]
	\centerline{\includegraphics[width=3.5in]{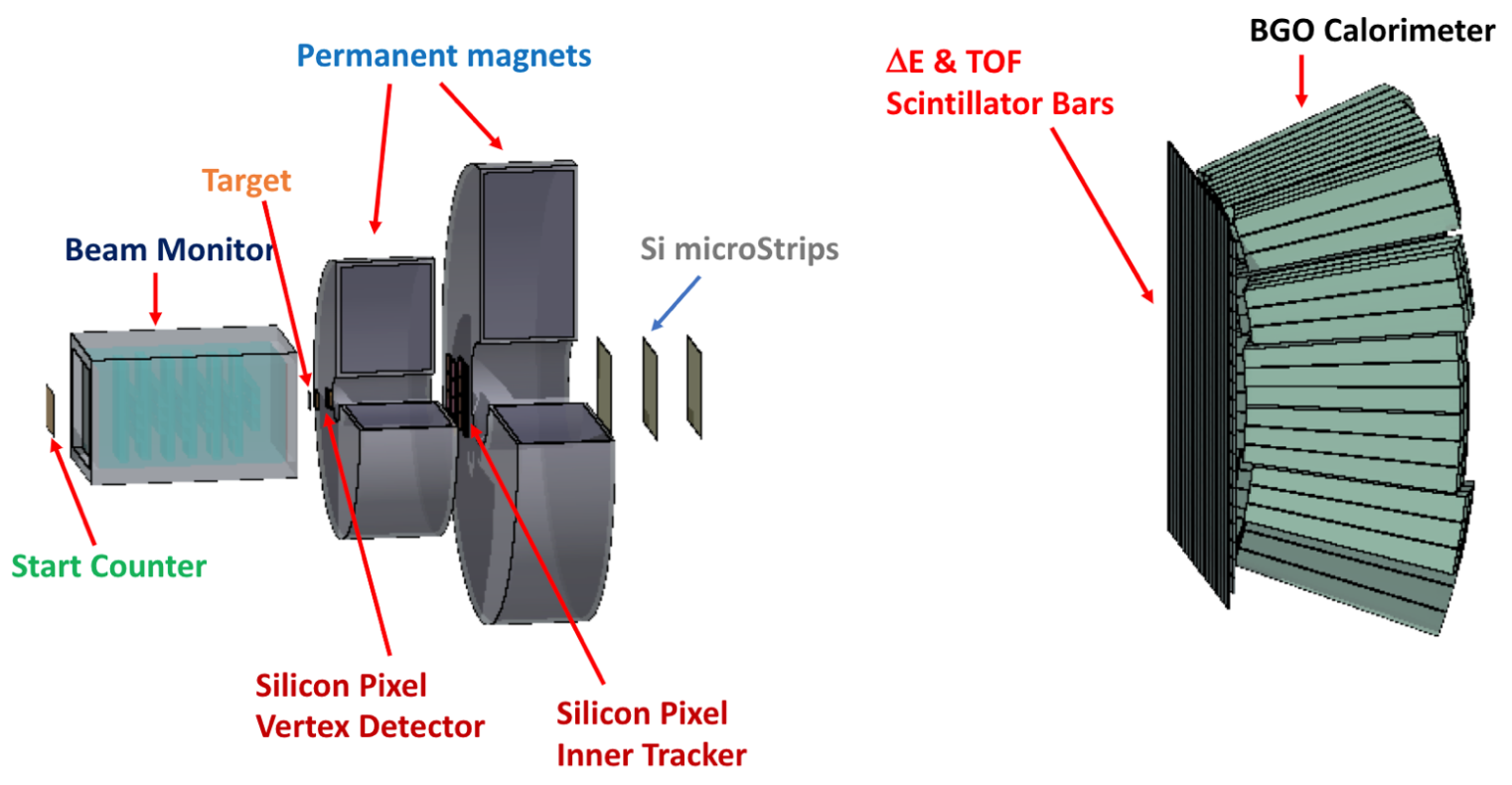}}
	\caption{Schematic view of the electronic setup with all the regions: pre-target, tracking and downstream.}
	\label{fig_electronic}
\end{figure}

The electronic setup of the FOOT detector \cite{b5} consists of a pre-target region, a tracking region and a downstream region and is devoted to the measurements of fragments with Z$\geq$4. Fig. \ref{fig_electronic} shows the entire experimental setup of this configuration. Three main regions can be identified in the experimental setup:
\begin{enumerate}
	\item The pre-target region composed by the Start Counter (SC) and the Beam Monitor (BM);
	\item the tracking region (including the interaction region as well) composed by the target followed by three stations of silicon pixels and strips detectors allocated upstream, between and downstream of two permanent magnets;
	\item the dowstream region composed by two orthogonal planes of thin plastic scintillator bars (TOF Wall, TW), placed at least 1 m away from the target, and a BGO calorimiter placed immediately after the TW.
\end{enumerate}

In order to achieve the required precision on the final cross-section measurements, the following benchmarks in resolutions need to be obtained with this setup:
\begin{enumerate}
	\item $\sigma (p)/p \sim$ 4--5\%;
	\item $\sigma (TOF) \sim$ 100 ps;
	\item $\sigma (E_{\mathrm{kin}})/E_{\mathrm{kin}} \sim$ 1--2\%;
	\item $\sigma (\Delta E)/\Delta E \sim$ 5\%.
\end{enumerate}

\subsubsection{Pre-target region}
\label{subsubsec:pretarget}
The pre-target region aims to monitor the beam, providing its direction and the interaction point on the target, and to count the number of impinging ions. The amount of material in this region has been strongly reduced in order to minimize the out-of-target fragmentation, as well as multiple scattering of the beam. Thus, two detectors have been designed and developed to fulfil such requirements.

The Start Counter (SC) is composed by a thin squared foil of plastic scintillator 250 $\mu m$ thick, with an active surface of 5 $cm$ transverse size. The light produced in the scintillator is collected laterally by 48 SiPMs, 12 per side, bundled in 8 channels, each reading the series of 6 SiPMs. The readout and powering of this system is provided by the WaveDAQ system \cite{b6}, which samples signals at rates up to 5 Gsamples/s in a dynamic range of 1 V. 
The main role of the SC is multiple: providing the Minimum Bias trigger, measuring the incoming ion flux (with an efficiency higher than 99\%), providing the reference time for all the detectors and allowing the Time-Of-Flight (TOF) measurement in combination with the TOF-detector (see description at the end of this paragraph) of the magnetic spectrometer.

The Beam Monitor (BM) is a drift chamber consisting of 12 layers of wires, with three drift cells per layer. Planes with wires oriented along the $x$ and $y$ axes are alternated in such a way to recontruct the beam profile, main goal of this detector. The BM operated at $\simeq$ 0.9 bar with a 80/20\% gas mixture of Ar/CO$_2$, at a working point between 1850 V and 2200 V, depending on the beam. This detector is place between the SC and the target, in order to measure the direction and impinging point of the beam ions on the target, crucial to discard events in which the beam has fragmented in the SC producing one or more deviated tracks. Thus, an high precision alignement is required between the BM and the devices downstream the target. The high spatial resolution achieved (100 $\mu m$) is fundamental to measure the direction of the fragments in inverse kinematic with the required precision. Moreover, the capability of the BM to provide information about the beam spot size is crucial to monitor the low intensity beams used for the acqusitions, especially when the monitoring with the standard facility devices is not always feasible, for example in centers for patients treatment.

\subsubsection{Tracking region}
\label{subsubsec:tracking}
The entire tracking system of the FOOT experiment is composed by three stations allocated upstream, between and downstream of two permanent magnets. 

The first tracking station is the vertex detector (VTX) of the experiment, organized in 4 different pixel sensor layers of 2$\times$2 cm$^2$ transverse dimension, placed along the $z$ axis, with a distance that guarantees a geometrical acceptance of $\sim$40$^\circ$ for the emitted fragments from the target. In order to fulfill the requirements of a low material budget and a high precision and efficiency, the technology of the MIMOSA-28 (M28) Monolithic Active Pixel Sensors (MAPS) has been adopted for each layer of the VTX. All four M28 sensors are thinned to 50 $\mu m$, then the overall material budget for the entire Vertex tracker is 200 $\mu m$. The architecture of the M28 integrates a binary readout and a zero suppression logic in chip to reduce the amount of data transferred. The VTX readout is based on a DE10 board system housing an Intel System-on-Chip (SoC) FPGA (Cyclon V) with a dual-core Cortex-A9 CPU. The FPGA part is interfaced with the sensors and with the DAQ control (trigger, timestamping and busy signals) and the CPU is used to send data to the central DAQ via the 1 GbE connection. The high spatial resolution achieved with this detector (5 $\mu m$), combined with the information from the BM, provides an angular accuracy at the milliradiant level, minimizing also the multiple scattering effect thanks to the reduced material of both BM and VTX.

The magnetic system design follows the requirements of the portability of the apparatus and the momentum resolution of the measurement. Thus, a magnetic system in air composed of two magnets, in Halbach configuration, has been chosen, also because it allows an additional tracking station in the middle. In Halbach configuration, an approximately dipolar magnetic field is obtained in the internal hole of a cylindrical permanent magnet. The magnetic field increases with the external cylinder radius while decreases with the gap radius. So in order to match the final momentum resolution producing the needed ($B \times L$) and at the same time have an angular acceptance of 10$^\circ$ for the emitted fragments, two different magnet dimensions have been chosen. The first magnet with gap diameter of 5 cm while the second of 10.6 cm can provide respectively a maximum intensity of 1.4 T and 0.9 T along the $y$ axis in the internal cylindrical hole. Thanks to a detailed field map, this allows to reach the intrinsic achievable accuracy of about 10 $\mu m$.

The second tracking station, in between the two aforementioned magnets, is the Inner Tracker (ITR) and it is composed by two planes of pixel sensors to track the fragments in the magnetic region. Each plane covers a sensitive area of about 8$\times$8 cm$^2$, with 16 M28 sensors per layer and this choice was led by the fact that these sensors are expected not to be significantly affected by the foreseen residual magnetic field in between the two FOOT magnets, as well as the low material budget and the consequent reduction of multiple scattering and out-of-target fragmentation. The ITR is composed by four ladders, each made of two modules of M28-sensor layer glued on the opposite sides of a support structuree, made of low density silicon carbide (SiC) foam, 2 mm thick.

The last tracking station is a Microstrip Silicon Detector (MSD) composed by three $x$-$y$ planes 9.6$\times$9.3 cm$^2$ active area, separated one to each other by a 2 cm gap along the beam direction and positioned right after the second magnet. This will ensure the needed angular acceptance to measure ions with Z $>$2. In order to reduce the amount of material and to ensure the $x$-$y$ coordinate readout, two perpendicular Single-Sided Silicon Detector (SSSD) sensors thinned down to 150 $\mu m$ will be used for each MSD $x$-$y$ plane. A minimum strip pitch size of 50 $\mu m$ has been chosen in order to minimize fragment pile-up in the same strip. Each SSSD is readout by 10 VA1140 chips for a total of 640 channels. The front-end hybrids, hosting the readout chips, will be glued at one side of each silicon module minimizing the dead space in the beam region. A digital readout of strips with pitch of 150 $\mu m$ would provide a spatial resolution of $\simeq$ 40 $\mu m$, but with analog readout a further factor 3 could be easily gained \cite{b7}, with the additional advantage to measure also the dE/dx, for each $x$-$y$ view of each layer independently. The analog signals provided by the VA1140 readout chips are digitized by 1 MHz 12-bits ADC and their data are sent to a TERASIC DE10 nano board for data collection and event shipping to the general FOOT DAQ.

\subsubsection{Downstream region}
\label{subsubsec:downstream}
The downstream region is the last part of the apparatus, placed at least 1 m away from the target and has the main goal to provide the stop of the TOF and to measure both energy loss and fragment kinetic energy.

The TOF Wall (TW) is composed of two layers of 20 plastic scintillator bars each arranged orthogonally and wrapped with reflective aluminum and darkening black tape. Each bar is 0.3 cm thick, 2 cm wide and 44 cm long. The two orthogonal $x$-$y$ layers form a 40$\times$40 cm$^2$ active area detector that provides the measurements of the energy deposited $\Delta E$, the TOF, with the start from SC, and the hit position. The simultaneous measurement of the $\Delta E$ and the TOF provides the possibility to identify the charge Z of the impinging ions, fundamental for the mass identification and, together with the $x$-$y$ hit position, for the tracking of the fragments through the magnetic field.
Each of the two edges of the TW bars is coupled to 4 SiPM with 3$\times$3 mm$^2$ active area and 25 $\mu m$ microcell pitch. The signals of each channel (two channels per bar) are digitized at rates of 3-4 Gsamples/s, depending on the trigger scheme adopted, by the WaveDAQ system, the same readout shared also with the SC detector, as described in Section \ref{subsubsec:pretarget}. A total of 1024 samples are collected for each signal allowing to record the whole waveform, and to extract offline the time and the charge information. 
Also in this case, the FOOT requirements for the TOF resolution, discussed in Section \ref{subsec:electronic}, have been taken into account when chosing the thickness of the bars and the readout chain.

The FOOT calorimeter is the most downstream detector, designed to measure the fragments kinetic energy with the aim of identify their mass $A$. Depending on the energy of the incoming fragment, different processes can take place in the calorimeter in the energy range of interest for the FOOT experiment. It is also true that the highest performances are required for the case of target fragmentation, that in the inverse kinematic involves $^{12}$C and $^{16}$O up to 200 MeV/nucleon. At these energies, the main energy loss happens through electromagnetic interaction with the target electrons and nuclei and the best calorimeter performances can be reached. But there is also the probability for a fraction of the events that neutron production takes place in the calorimeter and part of the fragment energy escapes the calorimeter, causing a systematic error that spoils the energy resolution. However, the redundancy of the FOOT apparatus helps taking this phenomenon into account with their additional information.
BGO (Bi$_4$Ge$_e$O$_{12}$) crystals have been chosen according to the energy resolutions requirements for the final measurement: its high density ($\rho = 7.13$ g/cm$^{3}$) guarantees a high stopping power and, together with the light yield of $\simeq$ 10 photon/keV, allow to reach the requirements mentioned above.
Thus, the FOOT experiment is composed of 320 BGO crystals arranged in a disk-like geometry ($\simeq$ 20 cm radius) and divided in modules of 3$\times$3 crystals, in order to best handle the weight of the detector. Each BGO crystal is coupled to a 25 SiPMs matrix with an active surface of 2×2 cm$^2$, where each microcell has a pitch of 15 $\mu m$, small enough to have a linear response in the energy range up to about 10 GeV. The readout system equipping this detector is the same as for the TW, where the WaveDAQ system is used.

\subsection{Emulsion Chambers Setup}
\label{subsec:emulsion}
\begin{figure}[t]
	\centerline{\includegraphics[width=3.5in]{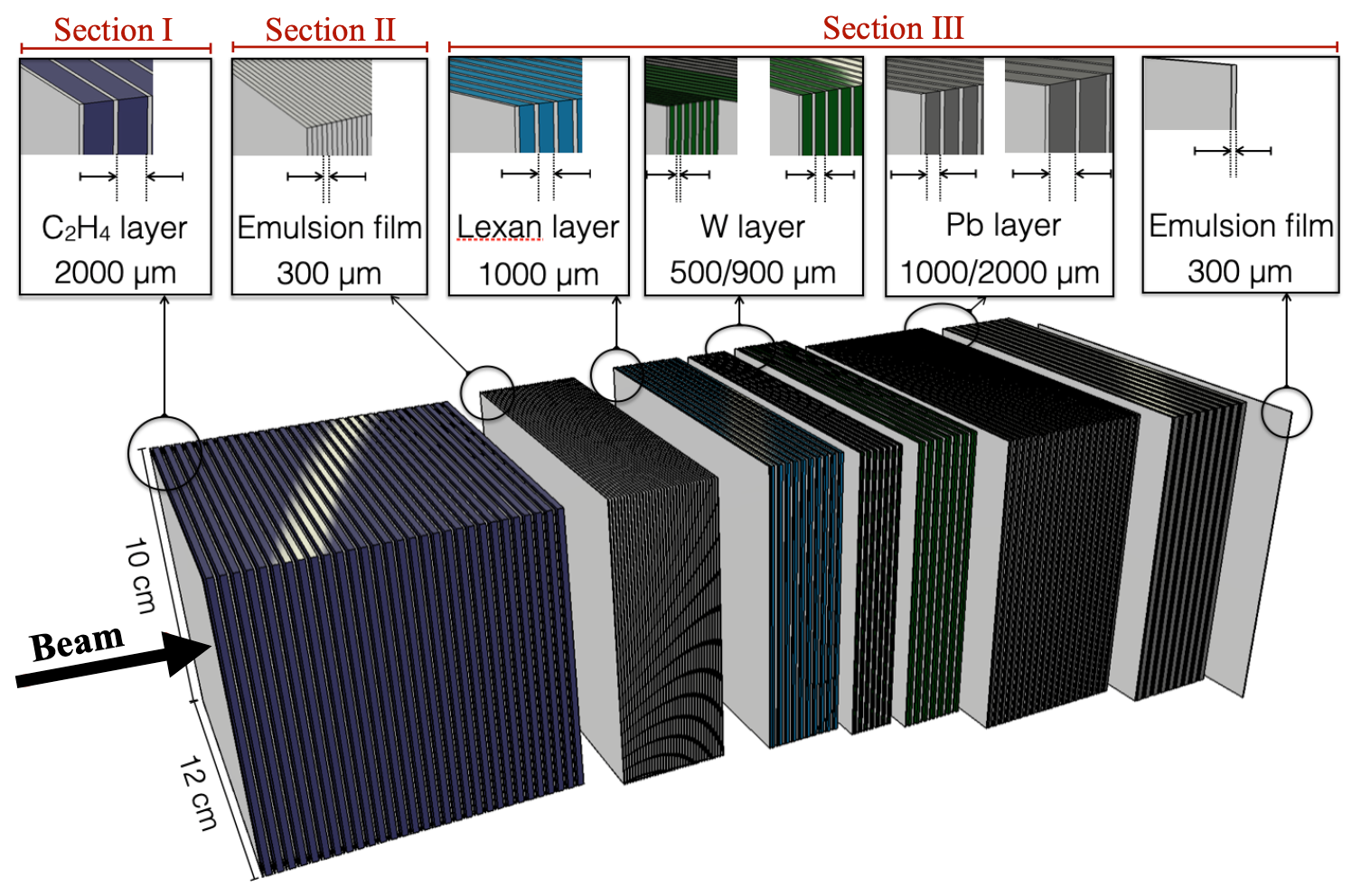}}
	\caption{Schematic view of the electronic setup with all the regions: pre-target, tracking and downstream.}
	\label{fig_emulsion}
\end{figure}

The FOOT experiment is equipped with an alternative setup as well, which includes an Emulsion Spectrometer (ES) and it is dedicated to the cross-section measurements of the lighter fragments \cite{b8}. Fig. \ref{fig_emulsion} shows the ES experimental setup.

The pre-target region is the same as the one used in the electronic setup, described in Section \ref{subsubsec:pretarget}, but both SC and BM are used only for beam monitoring purposes and their DAQ (described in Section \ref{sec:DAQsystem}) is completely separated with respect to the one used for the ES, acting as a complete and independent experiment.

The choice of this ES setup has been led by different aspects:
\begin{enumerate}
	\item nuclear emulsion detectors achieve the highest spatial resolution (sub-micrometric) for tracking ionizing particles;
	\item they integrate target and detector in a very compact setup (less than one meter long) and provide a very accurate reconstruction of the interactions (sub-millimetric resolutions) occurring inside the target;
	\item no power supply or any readout electronics is required, allowing to keep the ES compact and without spatial limitation.
\end{enumerate}
The possibility to measure particles emitted with an angle above 70$^\circ$, together with the very high spatial resolution and charge identification capability, made the nuclear emulsion technology an ideal choice for new generation of measurements of differential fragmentation cross sections.

In the FOOT experiment, the nuclear emulsion films consist of two 70 $\mu m$ thick sensitive layers placed on both sides of 120 $\mu m$ plastic base, resulting in a total thickness of 350 $\mu m$. The sensitive regions are made of a series of AgBr crystals of 0.2 $\mu m$ diameter scattered in a gelatine binder, capable to detect charged particles and to record their trajectories. 

The development of the films is a chemical process which enhances the latent images, including the growth of silver clusters (grains) with a diameter of 0.6 $\mu m$ which can be seen with an optical microscope. The density of grains is proportional to the charged particle ionization within the dynamical range. After the development, the emulsions are scanned by an automated system and the acquired images are, consequently, analyzed by a dedicated software to recognize clusters of dark pixels aligned, which represent the track produced by the penetrating particle.

The ES of the FOOT experiment has been designed with passive layers alternated to nuclear emulsions films and it is composed of three different sections each with a spefici purpose, as shown in Fig. \ref{fig_emulsion} :
\begin{enumerate}
	\item Interaction and vertexing region (ES Section 1), Paragraph \ref{subsubsec:section1};
	\item Charge identification region (ES Section 2), Paragraph \ref{subsubsec:section2};
	\item Momentum measurement region (ES Section 3), Paragraph \ref{subsubsec:section3}.
\end{enumerate}

\subsubsection{Interaction and vertexing region (Section 1)}
\label{subsubsec:section1}
The ES Section 1 is made of several elementary cells composed of layers of target element, Carbon or C$_2$H$_4$, alternated with emulsion films. When the ion beam interacts with the cells of this section, secondary fragments are emitted and detected by the following regions. The detector emulsion structure will track the fragments and reconstruct the interaction vertex position. The length of this section can be optimized for each data taking, taking into account ion beam, energy and target in order to achieve a statistically significant number of reactions.

\subsubsection{Charge identification region (Section 2)}
\label{subsubsec:section2}
The ES Section 2, aiming the charge identification for low Z fragments (H, He, Li), is made by elementary cells composed of four emulsion films. After the exposure and before the chemical development, four different thermal treatments were applied to the emulsions, hereafter denoted as R0 (not thermally treated), R1 (28$^\circ$C), R2 (34$^\circ$C) and R3 (36$^\circ$C). This procedure is needed because particles at the minimum of their ionizing power (MIPs) generate thin tracks whose grain density ranges from 30 to 50 grains/100 $\mu m$ and high ionizing particles cause a saturation effect spoiling the charge identification. In order to avoid it, the emulsions films are kept at a relatively high temperature (higher than 28$^\circ$C) and a high humidity (higher than 95\%) for about 24 hours. This technique will induce a fading which partially or totally erases the tracks of particles. Thus, films can be made unsensitive to MIPs and capable to identify charge of highly ionizing particles without the saturation effect.

\subsubsection{Momentum measurement region (Section 3)}
\label{subsubsec:section3}
The ES Section 3, devoted to the momentum measurement, is made of emulsion films interleaved with layers of passive material. As mentioned in Section \ref{subsec:emulsion}, the length, the number of passive layers and their thickness are set according to the incident beam energy. The materials used as passive layers are Lexan (C$_{16}$H$_{14}$O$_{3}$, 1.2 g/cm$^3$), tungsten (W, 19.25 g/cm$^3$) and lead (Pb, 11.34 g/cm$^3$).

\section{Trigger and Data Acquisition System}
\label{sec:DAQsystem}
\begin{figure}[t]
	\centerline{\includegraphics[width=3.5in]{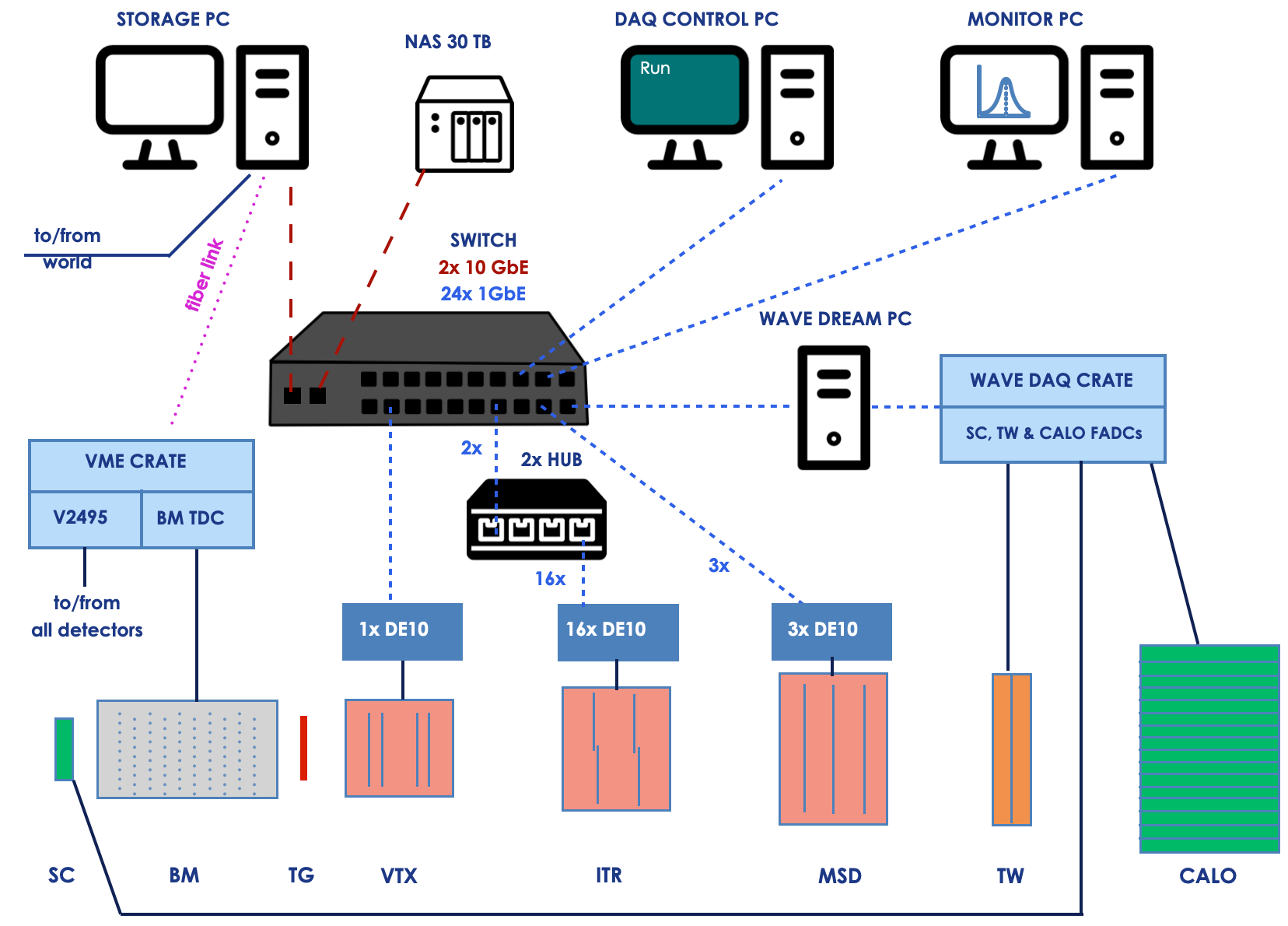}}
	\caption{Schematic view of the DAQ system of the FOOT experiment. The dashed lines represent the ethernet links (longdashed lines are 10 Gb/s links while dashed ones are for 1 Gb/s links) and the dotted lines represent the optical fiber links.}
	\label{fig_daq}
\end{figure}

The FOOT detector is equipped with a DAQ system, shown in Fig. \ref{fig_daq}, designed to acquire the largest sample size with high accuracy in a controlled and online-monitored environment. For a proficient system few guidelines are considered in the design: 

\begin{enumerate}
	\item the maximum acquisition rate should depend on the beam characteristics and/or on the slowest detectors in the experiment;
	\item the system should work in different laboratories and in different conditions;
	\item  it should allow an online monitoring of the data taking and a fast quality assessment of the acquired data;
	\item the storage system should be reliable and the data should be tansferred as soon as possible to the analysis center(s).
\end{enumerate}

The DAQ system that has been implemented for the whole apparatus is a flexible hierarchical distributed system based on linux PCs, SoC-FPGA boards, VME crates and boards and standard communication links like ethernet (dashed lines in Fig. \ref{fig_daq}) and optical fibers (dotted lines in Fig. \ref{fig_daq}). 
The control of the system is hosted on a PC used to run the DAQ GUI interface to start/stop a single run, to control and to configure other nodes in the system. Another PC (Storage PC) is used to collect the information coming from the different detectors, to perform an event building and to store on disk the acquired data. On the same PC, a MYSQL DataBase has the role to store the configuration data and to store DAQ process information. An electronic logbook interfaced with the DAQ system has been installed on the same machine. The actual readout systems can be in the form of standard VME boards placed in VME crates or in the form of PC or other boards reachable by the DAQ via ethernet. The ethernet link is required by the DAQ system for providing commands and receiving feedback. The data to be collected can use ethernet, USB or optical fibers to reach the Storage PC. In case of VME crates, two options are considered to dispatch DAQ commands on the different boards in the crate: a VME Bridge board using optical fiber connections or a Single Board Computer (SBC) in each crate. VME boards reading different detectors can be placed inside the same crate.

The trigger of the experiment can be generated with different conditions and it is distributed system-wide along with a redundant time-tagging mechanism that helps to correctly associate the generating trigger with the data fragments acquired. 
The main trigger of the experiment is the Minimum Bias trigger based on signals provided by the SC (Section \ref{subsubsec:pretarget}). Each SC signal is discriminated into the WaveDAQ system and the trigger is fired when the multiplicity of the channels above the thresholds exceeds a programmable value (majority trigger). This choice avoid any source of systematics due to the trigger selection. A fragmentation trigger asking for activity outside the central bars of the TW in a logical OR with a prescaled Minimum Bias trigger can also be used to enhance the fraction of recorded fragmentation events. The electronics that will be used to perform the trigger function is a CAEN V2495 board, whose FPGA and internal logic is fully programmable. The maximum acquisition rate in Minimum Bias would depend on the slowest detectors in the experiment: the MIMOSA 28 chips in the pixel tracker, which have a frame readout time of 185.6 $\mu s$, needed to readout about 10$^6$ pixels per chip. Thus, the overall maximum readout rate is fixed at about $R_{\mathrm{max}} =$ 5 kHz. The system is designed to handle a maximum DAQ rate of $R_{\mathrm{DAQ}} = R_{\mathrm{max}}$, but in order to reduce pile-up effects in the MIMOSA chips the actual trigger rate will be of the order of $R_{\mathrm{trigger}} =$ 1 kHz. With this rate, considering a duty cycle of $f_{\mathrm{dc}} =$ 30\%, during stable running conditions, up to $N_{\mathrm{day}} \simeq$ 86400 $\cdot$ 1k $\cdot$ 0.3 $=$ 26 M events per day can be collected with a Minimum Bias trigger.

As shown in Fig. \ref{fig_daq}, the steering of the acquisition process and the reading of the other nodes is managed through an ethernet switch connected via a 10 Gb/s cable and a CAEN V2718, a VME to PCI Optical Link Bridge. The switch is used to collect all the data from the detectors connected via 1 Gb/s ethernet connection: the whole tracking system, based on 20 DE10-nano or DE10 Terasic boards, the Time Of Flight detectors system and the calorimeter, both based on the WaveDAQ boards. 
The DE10-nano boards have an FPGA for detector reading and a dual core ARM cortex 9 processor for event formatting, zero suppression and data shipping via ethernet. The WaveDAQ boards send data to an intermediate PC providing data calibration, compression and data shipping for both TOF and calorimeter detectors. The VME to PCI Optical Link Bridge in the storage PC is connected to a VME crate holding the trigger board V2495 and the Beam Monitor discriminators and TDC board CAEN V1190B.

\begin{figure}[t]
	\centerline{\includegraphics[width=3.5in]{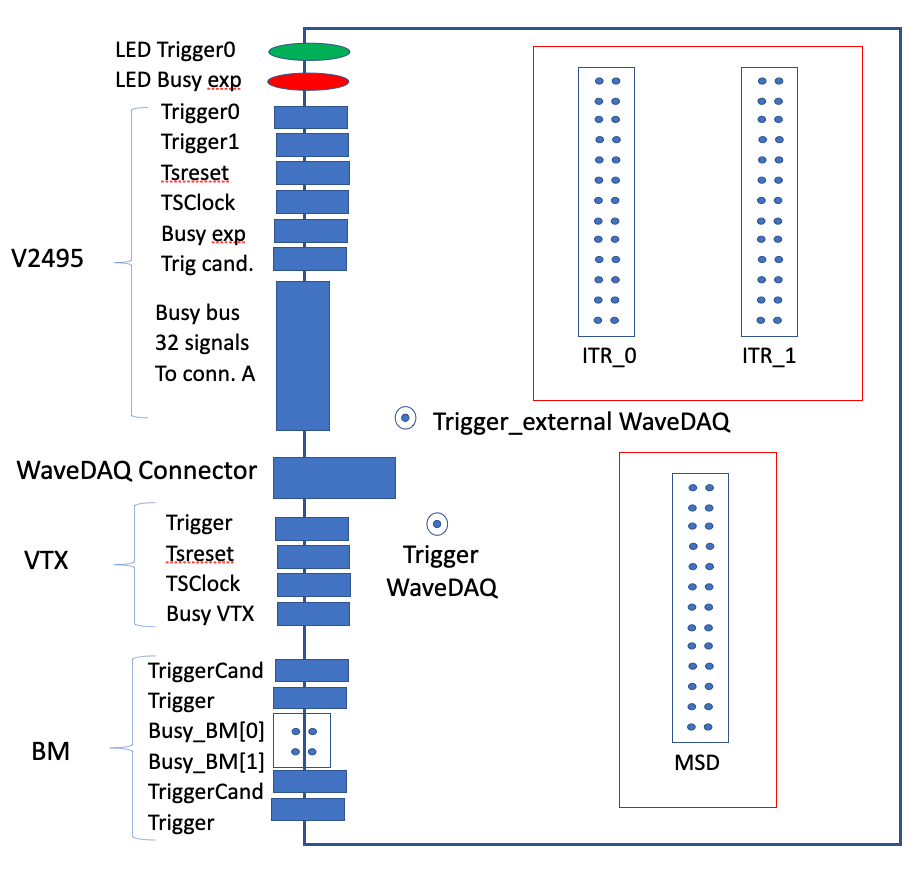}}
	\caption{Schematic view of the custom board (``patch panel''). The pins on the left are the ones in the front of the board, for signals from/to the trigger board V2495, the WaveDAQ system, and the VTX and BM readout boards. Connectors for the ITR and MSD detectors are also reported.}
	\label{fig_panel}
\end{figure}

Another important part of the DAQ system, not represented in Fig. \ref{fig_daq}, is a custom board (called ``patch panel''), schematize in Fig. \ref{fig_panel}. The main goal of this board answers to the necessity to synchronize all the subdetectors of the FOOT experiment and to handle the trigger, in order to be distributed system-wide. 

This custom board is necessary since the DAQ system must handle different types of signals which are designed for different subdetectors necessities, thus, a single and specific board capable of handling and distributing accordingly all the input and output signals is crucial for the performances required to the DAQ system of the FOOT experiment. Moreover the patch panel size has been designed like a VME 6U board, in order to be placed into the VME crate already used for the trigger board V2495, the TDC boards and discriminators for the BM. This configuration allows to steer the power of the patch panel directly from the crate itself. The board has been equipped with connectors specifically designed for the trigger board V2495 and for each subdetector: Beam Monitor (BM), Vertex (VTX), Inner Tracker (ITR) and MSD (Micro Strip Detector). A dedicated connector to send and receive the signals from the WaveDAQ system for the Start Counter (SC), the TOF Wall and the calorimeter. Two additional pins have been included: one to receive and handle the trigger signal coming from the WaveDAQ system (Trigger$\_$external WaveDAQ) and the other one to distribute this trigger signal to all the other subdetectors which are connected to the patch panel (Trigger WaveDAQ).

Taking into account the possibility to perform the zero suppression algorithms online on most of the detectors in the front-end electronics, the estimated average event data size is of the order of 100 kB to be acquired at an average rate of 1 kHz. This fixes the data collection capability of the system at the busiest node (the storage PC) to be of the order 60 MB/s on average. Taking into consideration a safety factor of 4, a minimal bandwidth of 240 MB/s is considered in the DAQ system design: the storage PC is connected with central switch network via a 10 Gb/s ethernet link. 
The availability of RAM memories along the data collection paths (in the FPGAs, in the DE10, in the PCs, in the switch and in the CAEN boards) allows an almost complete decoupling of the trigger signal time from the event building time in the storage PC that can be several seconds apart, while still keeping an average DAQ rate of 1 kHz (with rate peaks of 5 kHz). Moreover, several buffers and pipelines in the systems are used to reduce at minimum the dead-time and the data losses.

Since the system has been designed as mentioned before, the data rate exceeds the average performance of standard hard disks. For this reason, the whole system is designed to store data on a SSD disk (mean data transfer rate 400 MB/s), placed in the storage PC for temporary storage during data taking and to transfer data to a dedicated $>$20 TB) NAS system for storage during idle times.

The data collected are processed in real time for quality assessment. Several sets of online monitoring information are available:
\begin{enumerate}
	\item simple information pieces, like counters or rates, are coming from each subdetector
	\item a second more informative information come in the form of histograms filled on each PC in the system using local data, to show detector occupancies, particle arrival times, particle energies, collected charges and so on;
	\item a third and more powerful online information comes from a fast online event reconstruction performed on the fly on a fraction of events, in order to have track momentum spectra, TOF, A and Z reconstructed for charged tracks.
\end{enumerate}

All these data are available to the DAQ crew and detector experts during data taking, using an online monitoring system able to be distributed on several PCs in parallel.

\section{Conclusions}
The FOOT (FragmentatiOn Of Target) experiment has been designed to perform measurements of differential cross sections for the production of charged fragments in the nuclear interaction between ion beams (p, $^4$He, $^{12}$C, $^{16}$O) and targets (H, C, O) of interest for charged Particle Therapy (PT) and space radioprotection. For the PT, an inverse kinematic approach is used to measure the cross sections for the production of charged fragments in p+C and p+O collisions in the energy range 50-200 MeV/nucleon, using beams of $^{12}$C, $^{16}$O on graphite and polyethylene targets. For the radio protection in space, the same apparatus is used to investigate the double differential cross sections of the projectile fragmentation process for beams of 4He, $^{12}$C, and $^{16}$O impinging on graphite, polyethylene and PMMA targets up to 500 MeV/nucleon for charged PT and up to 800 Mev/nucleon for space radioprotection.

The FOOT detector takes advantage from two different and complementary setups a magnetic spectrometer and an emulsion spectrometer, for the tracking and the identification of the fragments in order to provide high acceptance, efficiency and identification capability in a wide dynamical range that includes protons and heavier ions (up to $^{16}$O).

The construction of the detector is being finalized and its DAQ system (involving both the collection of data and the on-line monitoring of the data quality) is evolving along with the detector development and assembly. Several beam tests have been already performed in different treatment or experiment rooms. All of them have been essential opportunities to validate and further improve the performances of each subdetector.

The experiment started its scientific program using the Emulsion setup at GSI (in Darmstadt, Germany), in 2019 with $^{16}$O ions at 200 and 400 MeV/nucleon on C and C$_2$H$_4$ targets, and in 2020 with $^{12}$C ions at 700 MeV/nucleon, on the same targets. Data have been analized and almost the 99\% of the charge has been identified for the reconstructed charged particles.
The Electronic setup is under construction and a first data taking in this configuration is being scheduled at CNAO, using $^{12}$C ions at 200 MeV/nucleon.

An upgrade of the FOOT experiment is being already discussed and involve the evaluation of the neutron production together with the charged fragments. This evaluation plays a crucial role in constraining more strongly Monte Carlo nuclear production models that are relevant both for Particle Therapy and radioprotection in deep space. Thus, several studies on providing neutron detection capability in the FOOT experiment are currently ongoing.


\end{document}